\documentclass[prb,twocolumn,superscriptaddress,showpacs,showkeys]{revtex4}

\usepackage{graphics}
\usepackage{graphicx}

\begin{document}

\newcommand{\RR}{\mathrm{\mathbf{R}}}
\newcommand{\rr}{\mathrm{\mathbf{r}}}
\newcommand{\defin}{\stackrel{def}{=}}

\bibliographystyle{apsrev}

\title{Effect of post-growth annealing on the optical properties of InAs/GaAs quantum dots:
A tight-binding study}

\author{R. Santoprete}
\affiliation{Instituto de F\'{\i}sica, Universidade Federal do Rio de
Janeiro, Caixa Postal 68528, 21941-972 Rio de Janeiro, Brazil}
\affiliation{Laboratoire de Physique des Interfaces et des Couches Minces, CNRS-Ecole Polytechnique, 91128 Palaiseau cedex, France}
\author{P. Kratzer}
\affiliation{Fritz-Haber-Institut der Max-Planck-Gesellschaft, Faradayweg 4-6, 14195 Berlin, Germany}
\author{M. Scheffler}
\affiliation{Fritz-Haber-Institut der Max-Planck-Gesellschaft, Faradayweg 4-6, 14195 Berlin, Germany}
\author{R. B. Capaz}
\author{Belita Koiller}
\affiliation{Instituto de F\'{\i}sica, Universidade Federal do Rio de
Janeiro, Caixa Postal 68528, 21941-972 Rio de Janeiro, Brazil}

\date{\today}

\begin{abstract}

We present an atomistic study of the strain field, the one-particle electronic spectrum and the oscillator strength of the fundamental optical transition in  chemically disordered In$_{x}$Ga$_{1-x}$As pyramidal quantum dots (QDs).
Interdiffusion across the interfaces of an originally ``pure'' InAs dot  buried in a GaAs matrix is simulated through a simple model, leading to atomic configurations where the abrupt hetero-interfaces are replaced by a spatially inhomogeneous composition profile $x$. 
Structural relaxation and the strain field calculations are performed through the Keating valence force field (VFF) model, while the electronic and optical properties are determined within the empirical tight-binding (ETB) approach.
We analyze the relative impact of two different aspects of the chemical disorder, namely: (i) the effect of the strain relief inside the QD, and (ii) the purely chemical effect due to the group-III atomic species interdiffusion.
We find that these effects may be quantitatively comparable, significantly affecting the electronic and optical properties of the dot. Our results are discussed in comparison with recent luminescence studies of intermixed QDs.   
\end{abstract}



\maketitle

\pagebreak

\section{Introduction}

Nanometer-size semiconductor quantum dots (QDs) have been the subject of many studies in the past years, due to their
potential applications in optoelectronic devices and to their peculiar physical properties.\cite{bimberg1}
As one
particularly attractive feature, they offer the possibility to
tailor the character of the QD electron (or hole) energy levels
and of the energy of the fundamental optical transition by
controlling the size, shape and composition of the QD through the
growth process.
Experimentally, InAs QDs in GaAs have been grown both by molecular beam epitaxy and metal-organic chemical
vapor deposition.
In most growth processes, nonuniform Ga incorporation in nominally InAs QDs has been 
reported.\cite{scheerschmidt,joyce15981,kegel1694,xu3335,fafard2374,lita2797,
garcia2014,rosenauer3868,bruls1708,chu2355,fry,jayavel1820,joyce1000,lipinski1789,zhi604}
Photoluminescence studies of annealed QDs have shown a blue-shift of their emission line,
\cite{leon1888,malik1987,lobo2850,xu3335,fafard2374}
%
which was suggested to reflect diffusion of Ga atoms from the
matrix material into the QD during annealing. However, it is not
clear to which extent the blue-shift is a consequence of chemical
substitution (bulk GaAs has a wider band gap than InAs), and to
which extent it is due to reduced strain in the QD after Ga
interdiffusion, which also causes a band-gap widening.  The
recently observed change in the photoluminescence polarization
anisotropy upon annealing~\cite{ochoa192} represents a further
interesting but not yet fully understood QD property.

From a theoretical point of view, a realistic treatment of elastic, electronic 
and optical properties of such heterostructures must consider a non-uniform In$_{x}$Ga$_{1-x}$As composition 
profile inside the QD, which we refer here as chemical disorder.
Several theoretical works deal with chemical disorder, either from a macroscopic continuum approach, or within a microscopic model. 
Microscopic models provide an atomistic treatment, as required for a more reliable description of disordered 
heterostructures, taking into account the underlying zinc-blende structure and thus the correct $C_{2v}$ symmetry of pyramidal QDs.\cite{pryor1}
For the elastic properties, previously adopted macroscopic approaches involve a finite element 
analysis\cite{stoleru131,liao} or a Green's function method,\cite{califano389}
both in the framework of the continuum elasticity theory.
Microscopic approaches rely on empiric interatomic potentials, such as the Tersoff type,\cite{tersoff5566} adopted for 
truncated pyramidal QDs,\cite{migliorato115316} and the Keating\cite{keating,martin} valence force field (VFF) model, used in the study of truncated conical QDs.\cite{shumway125302}

A physical aspect indissociable from atomic interdiffusion is the strain relief mechanism due to the presence of chemical disorder, an effect that has not been highlighted by the previous theoretical studies.
We study here square-based pyramidal In$_{x}$Ga$_{1-x}$As QDs within a combination of VFF and empirical tight-binding (ETB) models, where 
we distinguish between two different aspects of the chemical disorder on the electronic and optical properties, 
namely the effect of the strain relief inside and around the QD and the purely chemical effect due to the presence of new atomic species (Ga atoms) penetrating inside the QD.

From the structural point of view, we calculate the strain field inside and around the dot and directly compare these results with those from a pure InAs/GaAs QD of the same size and geometry. 
This allows a quantitative analysis of the strain relief mechanism due to alloying.
To simulate the chemical disorder, we employ an atomistic diffusion model, where the degree of interdiffusion (and thus the degree of chemical disorder) can be controlled, so that a direct comparison between a chemically pure InAs/GaAs QD and chemically disordered In$_{x}$Ga$_{1-x}$As dots can be made.

Regarding the electronic properties, previous studies relied on macroscopic approaches such as 
the single band effective mass approximation\cite{barker13840,fry,roy235308,vasanelli551} or 
the multiband $\mathbf{\mathrm{k}} \cdot \mathbf{\mathrm{p}}$ model,\cite{heinrichsdorff:98,park144,sheng125308,sheng394,stoleru131} 
or on microscopic approaches as the empirical pseudopotential model\cite{bester:073309,bester161306,bester47401,shumway125302} 
or the empirical tight-binding (ETB) model.\cite{klimeck601}
The macroscopic models, working with envelope wavefunctions,
are applicable to smooth composition gradings only~\cite{sheng125308,gunawan:05} 
and cannot properly address the effect of microscopic composition
fluctuations, which are characteristic of annealed samples.
We show here that, within ETB, it is possible to examine separately
how two different aspects of chemical disorder affect  the QD electronic and optical properties, namely the effect of the strain relief inside the QD 
and the purely chemical effect due to In $\leftrightarrow$ Ga  interdiffusion. 
We decouple these effects by performing two independent calculations of the single particle electronic 
bound states and the fundamental optical transition: One in a ``physical'' (strained) QD, 
and the other in an artificially strain-unaffected QD, where only chemical disorder effects play a role.
Piezoelectric effects were not included here, since they become important only for larger
QDs.\cite{bester:045318}

This paper is organized as follows: 
In Sec. II we present the diffusion model employed to simulate the
chemical disorder, and we outline the procedure for the
calculation of the electronic and optical properties within the ETB model.
In Sec. III we present our results, and in Sec. IV a summary and conclusions.

\section{Formalism}

\subsection{Structural properties}

We start 
with a square-based pyramidal InAs QD with \{101\} facets and a one-monolayer thick InAs 
wetting layer, all embedded in a GaAs matrix.
We restrict ourselves for the present purpose to this simple QD
shape since the relation between the blue-shift and degree of
interdiffusion was found to be only weakly
shape-dependent.\cite{gunawan:05}
The pyramid base is 6.8 nm, the height is 3.4 nm, and the external dimensions of the GaAs matrix are 
$25a \times 25a \times 17.067a$, where $a=5.653$ \AA \ is the lattice constant of bulk GaAs.
The system contains 85000 atoms, and periodic boundary conditions are applied.
Chemical disorder is introduced in the system by allowing the interdiffusion of 
Ga and In atoms across the QD boundaries.
Since the anion sublattice occupation is not affected by disorder, we discuss the model in terms of the 
group-III species fcc sublattice.  
Interdiffusion is modeled atomistically, i.e., each
In atom may exchange its position with one of its Ga nearest
neighbors according to a probability $p$ proportional to the
concentration of Ga atoms around the In atom ($p=N_{\rm Ga}/12$,
where $N_{\rm Ga}$ is the number of Ga atoms among its 12 fcc
nearest neighbors). If an exchange takes place, the affected Ga
atom is picked randomly among the Ga nearest neighbors.
We stress that the microscopic rules employed to model diffusion
are compatible with Fick's law of chemical diffusion on the
macroscopic level. In our diffusion model, one era of duration
$\Delta t$ is completed after all cations in the system have been
attempted to move.
%
The interdiffusion process is iterated for a discrete number
$\tau$ of eras, and the resulting final atomic configuration at
$t=\tau \Delta t$ defines the QD to be analyzed. The parameter
$\tau$ quantifies the extent of alloying in the system, and simulates the 
anneal temperature in controlled intermixing experiments.\cite{fafard2374}
In order to give
some insight about the overall behavior to be expected from our
assumptions, we present initially a description for the evolution of
the {\it average} occupation probabilities at each site.
More explicitly, we call $P_{\rm In} (\RR_{i}, t)$ the probability of having an In atom in a cation 
lattice site at the position $\RR_{i}$ at a given time step $t$ ($t = 0,1,2,3,\ldots, \tau$).
This probability defines the average local concentration $x$ of In atoms. 
Obviously, the probability of having a Ga atom at the same position and at the same time step is 
$P_{\rm Ga} (\RR_{i}, t) = 1- P_{\rm In} (\RR_{i}, t)$.
The average spatial and temporal evolution of $P_{\rm In} (\RR_{i}, t)$ is described by the equation

\begin{eqnarray}
\label{strain:eq_diffusione}
P_{\rm In} (\RR_{i}, t) &=& P_{\rm In} (\RR_{i}, t -1)  \\ 
   &+& \frac{1}{12}\ P_{\rm Ga} (\RR_{i}, t -1) \cdot \sum_{j=1}^{12} P_{\rm In} (\RR_{i} + \vec{\xi}_{j}, t -1) \nonumber \\
   &-& \frac{1}{12}\ P_{\rm In} (\RR_{i}, t -1) \cdot \sum_{j=1}^{12} P_{\rm Ga} (\RR_{i} + \vec{\xi}_{j}, t -1), \nonumber 
\end{eqnarray}
where $\vec{\xi}_{j}$ is the $j$-th nearest neighbor position-vector in the fcc sublattice.
The following points should be mentioned:

\begin{enumerate}   
\item In and Ga atoms are treated symmetrically, thus the evolution of $P_{\rm Ga} (\RR_{i}, t)$ is given
by an equation analogous to (\ref{strain:eq_diffusione}), where the roles of In and Ga are interchanged.
It follows that the diffusion of In atoms into a GaAs-rich region proceeds identically to the diffusion of 
Ga atoms into an InAs-rich region.
\item Ga (In) atoms can penetrate at most $\tau$ lattice constants into
the QD (into the matrix), and $\tau = 0$ corresponds to no
interdiffusion taking place. 
\item The global concentration does not vary, i.e., the total number of cations of each species (In or Ga) in the 
system remains constant.
\end{enumerate}

A VFF model, parameterized as
described in Refs.~\onlinecite{pryor1,santoprete}, is then applied
to determine the atomic relaxations that minimize the total
elastic energy for the given distribution of species.
In the minimization process, each atom is moved along the direction of the force acting on it, and the
procedure is iterated until the force in each atom is smaller than $10^{-3}$eV/\AA.

\subsection{Electronic and optical properties}

The electronic and optical properties are studied within an ETB method, adopting a $sp^{3}s^{*}$ 
parametrization with interactions up to second nearest neighbors and spin-orbit coupling.\cite{boykin}
Strain effects are included by considering both bond length and bond angle deviations from ideal bulk InAs and 
GaAs.\cite{santoprete}
Bond length deviations with respect to the bulk equilibrium distances $d^{0}_{ij}$ affect the ETB Hamiltonian 
off-diagonal elements $V_{kl}$ as

\begin{equation}
V_{kl} \left( \left| \mathbf{R}_{\mathrm{i}}-\mathbf{R}_{\mathrm{j}}\right| \right) = V_{kl}(d_{ij}^{0}) \ \left( \frac{ d_{ij}^{0} }{ \left| \mathbf{R}_{\mathrm{i}}-\mathbf{R}_{\mathrm{j}} \right| } \right)^{n},
\label{scaling}
\end{equation}
where $\left| \mathbf{R}_{\mathrm{i}}-\mathbf{R}_{\mathrm{j}} \right|$ is the actual bond-length and $V_{kl}(d_{ij}^{0})$ 
is the bulk matrix element as given in Ref.~\onlinecite{boykin} ($k$ and $l$ label the different matrix elements).
The exponent $n$ is a parameter determined to reproduce the volume deformation potentials of InAs and GaAs, whose value 
was previously determined\cite{santoprete} as $n=3.40$ for all $k$ and $l$.
Strain effects may be easily removed from the ETB Hamiltonian.
The effects of the bond length deformations are completely removed from the Hamiltonian by taking  
$n = 0$ in Eq.~(\ref{scaling}). 
An equivalent transformation causes the effect of bond angle deviations from the ideal tetrahedral angles to be eliminated from the ETB Hamiltonian.
Single bound hole states $|h \rangle$ and electron bound
bound states $|e \rangle$ are calculated as eigenvectors of the
ETB Hamiltonian, using the folded spectrum
method.\cite{capaz,wangwang:94}
The optical transitions in the QD, treated within the electric dipole approximation, 
are quantified in terms of the dimensionless oscillator strength
\begin{equation}
\label{oscillator}
f_{eh}\ =\ \frac{2 |\langle e |\mathbf{p} \cdot \mathbf{\hat{e}}| h \rangle |^{2}}{m \hbar \omega_{eh}},
\end{equation}
where $|h \rangle$ is the initial QD hole bound state, $|e \rangle$ is the final QD electron bound state, 
$\hbar \omega_{eh}$ is the transition energy, $m$ is the free electron mass,
and $\hat{e}$ is the polarization unit-vector.
Within ETB, the electron and hole states are given by 
\begin{eqnarray}
|h \rangle &=& \sum_{\alpha \sigma \RR} C^{(h)}_{\alpha \sigma \RR} \ |\alpha \sigma \RR \rangle \nonumber \\
|e \rangle &=& \sum_{\alpha' \sigma' \RR'} C^{(e)}_{\alpha' \sigma' \RR'} \ |\alpha' \sigma' \RR' \rangle~,
\end{eqnarray}
and the electric dipole transition matrix element $\langle e |\mathbf{p} \cdot \mathbf{\hat{e}} | h \rangle$ can be approximately written as \cite{koiller4170}
\begin{eqnarray}
\label{dipolemoment}
\langle e |\mathbf{p} \cdot \mathbf{\hat{e}} |h \rangle \ & \cong \ &
\frac{i m}{\hbar}\ \sum_{\alpha' \sigma' \RR'} \sum_{\alpha \sigma \RR}
              C^{(e)\ *}_{\alpha' \sigma' \RR'}\ C^{(h)}_{\alpha \sigma \RR} \nonumber \\
   &\times&   \langle \alpha' \sigma' \RR' | H | \alpha \sigma \RR \rangle\ (\RR' - \RR) \cdot \mathbf{\hat{e}}\ ,
\end{eqnarray}
where $|\alpha \sigma \RR \rangle$ represents a general ETB basis vector  
($\alpha$ runs over the $s, p_{x}, p_{y}, p_{z}$ and $s^{*}{\rm~type}$ ETB orbitals, 
$\sigma$ labels the spins, $\RR$ the atomic sites), and 
$C^{(h)}_{\alpha \sigma \RR}$ and $C^{(e)}_{\alpha' \sigma' \RR'}$ are the expansion coefficients of the hole 
and electron QD bound states in the ETB basis. 

Expression~(\ref{dipolemoment}) can be easily evaluated, since it involves all known quantities.
Similarly to the electronic properties, for the optical properties the strain effects may also be 
completely removed from the calculation.
This is easily done by using in Eq.~(\ref{dipolemoment}) the strain-unaffected ETB Hamiltonian matrix elements for 
$\langle \alpha' \sigma' \RR' | H | \alpha \sigma \RR \rangle$, the strain-unaffected
wave function expansion coefficients for $C^{(h)}_{\alpha \sigma \RR}$ and $C^{(e)}_{\alpha' \sigma' \RR'}$,
and the ideal (bulk) zinc-blende interatomic vectors ($\RR - \RR'$).

\section{Results}

\subsection{Strain field}

\begin{figure*}
\begin{center}
\resizebox{140mm}{!}{\includegraphics{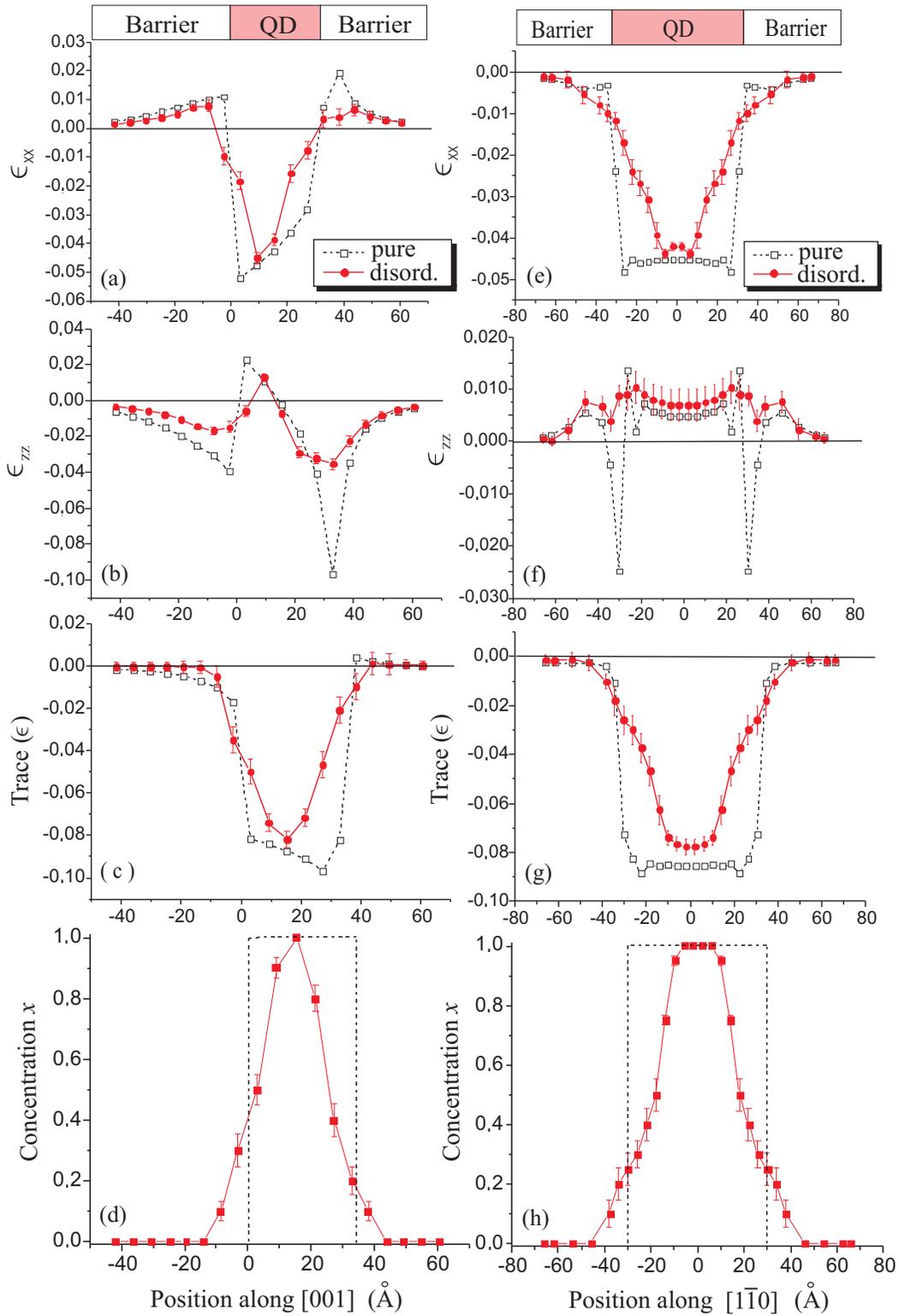}}
\caption{\label{strain_concentr_profiles}(Color online) Comparison between the components of the local strain tensor [panels (a)-(c) and (e)-(g)] and the concentration $x$ of In atoms [panels (d) and (h)] in a QD of pure InAs (Pure) and in a chemically disordered QD (Disordered) with $\tau =6$.
The panels on the left side show results calculated along a line oriented along the [001] 
direction and passing through the tip of the pyramid (the value 0 in the horizontal axis corresponds to the position of the wetting layer). 
The panels on the right side show results calculated along a line oriented along the 
[1$\bar{1}$0] direction and intersecting the [001]-oriented pyramid axis at height $h / 3$ from the base of the pyramid, where $h$ is the pyramid height.
The error bars indicate standard
deviations $\Delta \epsilon_{ij} = \sqrt{\langle \epsilon_{ij}^{2}
\rangle - \langle \epsilon_{ij} \rangle^{2}}/(N-1)$, where $N$=10.  }
\end{center}
\end{figure*}

Fig.~\ref{strain_concentr_profiles} shows a comparison of the average strain field and the local In average concentration between the chemically pure QD (corresponding to $\tau = 0$), given by the dotted lines, and the chemically disordered QD (chosen here with $\tau = 6$), given by the solid lines.
For the size of QD in this study, $\tau = 6$ allows for all but the 
innermost In atoms of the QD to diffuse out.
For disordered QD's, the results in Fig.~\ref{strain_concentr_profiles} for each property were obtained by averaging  over  those calculated for an ensemble of 10 different simulation supercells, all corresponding to $\tau = 6$, but generated from different sequences of random numbers at each interdiffusion step. 
In this way, the effect of composition fluctuations around the average values given in  
Eq.~(\ref{strain:eq_diffusione}) is reduced.  

The panels on the left side show the $xx$ component [panel (a)], the $zz$ component [panel (b)] and the trace [panel (c)]
of the local strain tensor, as well as the concentration $x$ of In atoms [panel (d)], along a line oriented along the [001] 
direction and passing through the tip of the pyramid.
The panels on the right side [(e) - (h)] show the corresponding quantities calculated along a line in the 
[1$\bar{1}$0] direction and intersecting the [001]-oriented pyramid axis at height $h / 3$ from the base of the pyramid, where $h$ is the pyramid height.
We observe from frames (d) and (h) that, according to our interdiffusion model, $\tau = 6$ corresponds to a penetration of the Ga atoms inside the QD (and consequently of the In atoms inside the GaAs matrix) of about 6 monolayers, i.e. about 17 \AA.
The error bars shown in the figure indicate standard deviations $\Delta \epsilon_{ij} = \sqrt{\langle \epsilon_{ij}^{2} \rangle - \langle \epsilon_{ij} \rangle^{2}}/(N-1)$, where $N$=10.
From the figure we may conclude that

\begin{enumerate}
\item Chemical disorder significantly reduces the absolute value of the strain field in the regions directly 
affected by the diffusion process, in agreement with experimental results.\cite{leon1888}
On the other hand, very small changes in the strain field occur in the regions not affected by interdiffusion, i.e. in the core of the pyramid and in the GaAs matrix, at large distances from the dot.
\item If interdiffusion takes place, the strain field varies more smoothly than in the case of a chemically pure QD. This is a direct consequence of the smooth variation of the concentration of In atoms across the heterointerfaces of the disordered dots.

\end{enumerate}

\subsection{Electronic and optical properties}

\begin{figure}
\begin{center}
\resizebox{85mm}{!}{\includegraphics{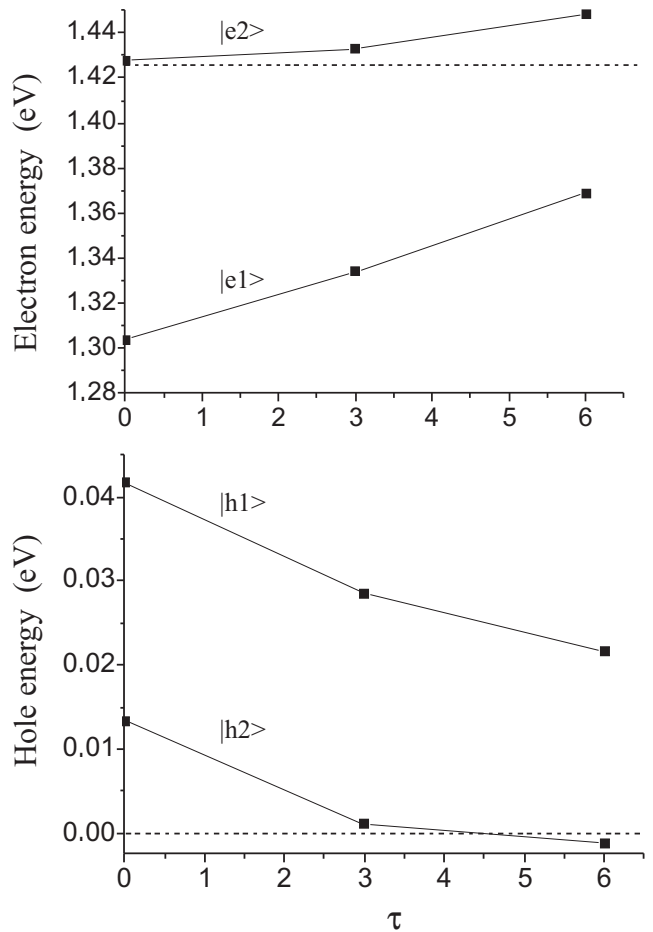}}
\caption{\label{energie_diff} First two QD bound electron (upper panel) and hole (lower panel) state energies as a function
of the degree of chemical disorder. The dashed horizontal lines represent the energies of the 
GaAs bulk conduction (upper panel) and valence (lower panel) band edges, delimiting approximately the energy range where a QD state is bound.}
\end{center}
\end{figure}

Fig.~\ref{energie_diff} shows the calculated eigenenergies of the QD bound states as a function of the degree of chemical disorder (characterized by the parameter $\tau$). 
The first two electron states ($|e1 \rangle$ and $|e2 \rangle$) are represented in the upper panel, 
while the first two hole states are shown in the lower panel.  A chemically pure QD corresponds to $\tau = 0$. 
The dashed horizontal lines represent the energies of the 
GaAs bulk conduction (upper panel) and valence (lower panel) band edges, delimiting approximately the energy range where a QD state is bound.
The figure shows that the electron state energies increase with increasing chemical disorder, while the hole state energies decrease, in agreement with previous empirical pseudopotential calculations\cite{shumway125302}.  
This behavior results in an increase of the frequency of the optical emission (blueshift), 
a phenomenon which has been experimentally observed.\cite{leon1888,lobo2850,malik1987,ochoa192}
The figure shows that, for $\tau = 6$, the QD gap is about 7\% larger than for $\tau = 0$.

\begin{figure}
\begin{center}
\resizebox{85mm}{!}{\includegraphics{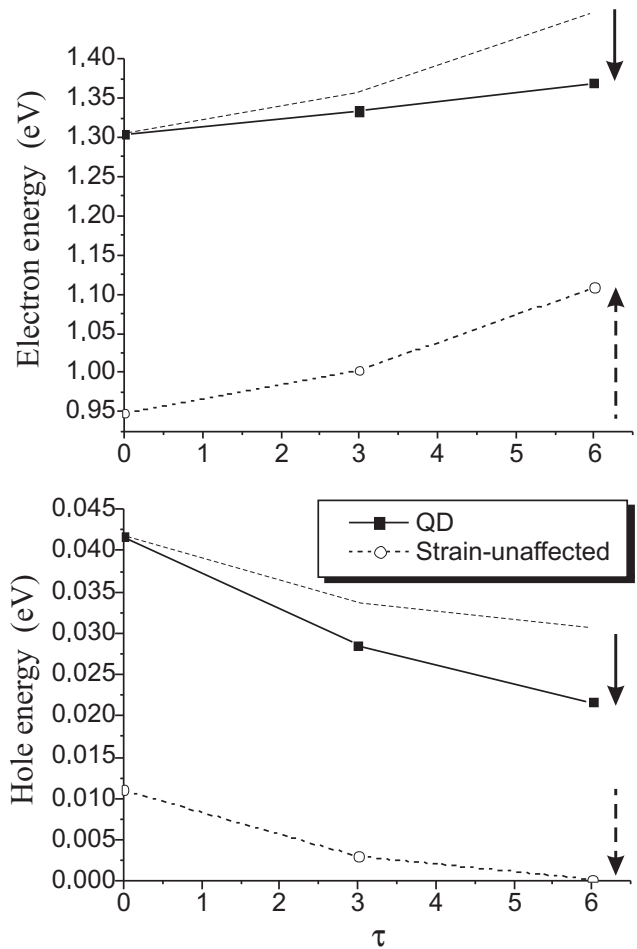}}
\caption{\label{energie_strain_bulk_diff}QD ground electron (upper panel) and hole (lower panel) state energy as
a function of the degree of chemical disorder. 
In each panel, we compare the results for the ``physical'' strained QD (QD) 
with those corresponding to the artificially strain-unaffected QD (Strain-unaffected).
In each panel, the uppermost dashed line is a guide for the eye parallel to the QD strain-unaffected energy curve, 
so that the strain relief contribution (represented by the solid arrow) can be visualized and
directly compared with the purely chemical effect of the disorder, represented by the dashed arrow.}
\end{center}
\end{figure}

Chemical disorder contributes to the results of Fig.~\ref{energie_diff} in two ways, namely by the strain
relief around the QD interfaces (see Fig.~\ref{strain_concentr_profiles}), and by the chemical effect due to the presence of Ga atoms inside the QD. 
These two effects can be decoupled by comparing the bound state energies of a strained QD with those of an
artificially strain-unaffected QD, as a function of the degree of disorder.
Fig.~\ref{energie_strain_bulk_diff} shows such comparison for the energy of the electron ground state $|e1 \rangle$ 
(upper panel) and of the hole ground state $|h1 \rangle$ (lower panel).
On each panel, the uppermost dashed line is a guide for the eye, parallel to the QD strain-unaffected energy curve and
starting from the $\tau = 0$ result for the physical QD. 
The strain relief contribution (represented by the solid arrow) can be 
directly compared with the purely chemical effect of the disorder, represented by the dashed arrow.
We see that these two effects are comparable, contributing in opposite directions for the electron state, and in the 
same direction for the hole state.
The purely chemical effect can be easily understood:
As the interdiffusion increases, the concentration $x$ of In atoms in the inhomogeneous alloy In$_{x}$Ga$_{1-x}$As inside the QD decreases.
The increase (decrease) of the electron (hole)  bound state energy as $x$ decreases is an alloying effect, so that the 
electron (hole) state energy tends (for $x \rightarrow 0$) to the bulk GaAs conduction band minimum (valence band maximum). 
Results in Fig.~\ref{energie_strain_bulk_diff} show that the chemical effects of disorder are partially canceled (enhanced) by the strain relief contribution for the electron (hole) state.

We now address the optical properties, focusing on the fundamental transition 
\mbox{$|h1 \rangle \rightarrow |e1 \rangle$}. 
In Table~\ref{results_table} we compare the results for a chemically pure QD (interdiffusion = off) with those for a 
chemically disordered QD (interdiffusion = on) with $\tau = 6$. 
For both cases, an additional comparison is made between a strained QD
(strain = on) and an artificially strain-unaffected QD (strain = off).
On the first two lines, we show the charge fraction 
$\displaystyle \Delta Q = \int_{QD}\ |\psi (\rr)|^{2} d^{3} r$ inside the QD for both the ground electron
state $|e1 \rangle$ and the hole ground state $|h1 \rangle$.
For the calculation of $\Delta Q$ in the chemically disordered case, the QD border
is taken the same as in the chemically pure case.
The third line shows the oscillator strength $f_{QD}$ of the transition 
$|h1\rangle \rightarrow |e1\rangle$ in the QD for unpolarized light, normalized to the oscillator 
strength $f_{\rm InAs}$ of the fundamental transition in bulk InAs for unpolarized light.
The fourth line gives the degree of anisotropy $I$ of the QD fundamental transition with respect 
to light polarization within the pyramid basal plane, defined as
\begin{equation}
\label{anisotropy}
I = \frac{|\langle e1 | \mathbf{p} \cdot \mathrm{\mathbf{\hat{e}}}_{+}  |h1 \rangle |^{2} - 
   |\langle e1 | \mathbf{p} \cdot \mathrm{\mathbf{\hat{e}}}_{-} |h1 \rangle |^{2}}
{|\langle e1 | \mathbf{p} \cdot \mathrm{\mathbf{\hat{e}}}_{+} |h1 \rangle |^{2} + 
     |\langle e1 | \mathbf{p} \cdot \mathrm{\mathbf{\hat{e}}}_{-} |h1 \rangle |^{2}},
\end{equation}
where $\mathrm{\mathbf{\hat{e}}}_{+}$ and $\mathrm{\mathbf{\hat{e}}}_{-}$ are unitary vectors along the 
inequivalent basal plane directions [110] and [1$\bar{1}$0] respectively.
The fifth line shows the oscillator strength $f_{[001]}$ of the QD fundamental transition for light linearly
polarized along the [001] direction, normalized with respect to $f_{QD}$.
Finally, the last line of Table~\ref{results_table} gives the relative change $(g-g_{0})/g_{0}$ of the optical QD gap $g$ with respect to the gap $g_{0}$ corresponding to the case ``strain off'' and ``interdiffusion off''. 
As for the case of the electronic properties, the direct comparison of the ``physical'' results with those of the 
disordered case and of the strain-unaffected case allows us to distinguish between the strain relief effect and the 
 chemical effect, both of which are due to chemical disorder.

\begin{table}
\begin{ruledtabular}
\begin{tabular}{ccccc} 
Strain & off & on & off & on \\ 
Interdiffusion & off & off & on & on \\ \hline
$\Delta Q_{|e1 \rangle}$ & 75\% & 64\% & 74\% & 34\% \\
$\Delta Q_{|h1 \rangle}$ & 11\% & 54\% & 11\% & 31\% \\
$f_{QD} / f_{InAs}$ & 12\% & 19\% & 10\% & 26\% \\
$I$ & $\sim 0$\footnote[1]{within our numerical precision}  & 2.5\% & $\sim 0$\footnotemark[1] & $(0.25 \pm 0.05)$\% \\
$f_{[001]} / f_{QD}$ & $\sim 0$\footnotemark[1] & 8\% & $\sim 0$\footnotemark[1] & $< 10^{-2}$\% \\
$(g-g_{0})/g_{0}$ & 0 & 35\% & 18\% & 44\% \\ 
\end{tabular}
\end{ruledtabular}
\caption{\label{results_table}Comparison of optical properties between different strain states [``strain on'' (= ``physical'' QD) or ``strain off''
 (= artificially strain-unaffected QD)] and different degrees of
interdiffusion  (``interdiffusion on'' (= chemically disordered QD) or ``interdiffusion off'' (chemically pure QD)).
The first two lines show the charge fractions within the QD corresponding to the ground electron ($\Delta Q_{|e1 \rangle}$) and
hole ($\Delta Q_{|h1 \rangle}$) state. 
The third line shows the oscillator strength $f_{QD}$ of the fundamental transition 
$|h1\rangle \rightarrow |e1\rangle$ in the QD, normalized with respect to the oscillator strength $f_{\rm InAs}$ of the 
fundamental transition in bulk InAs.
The fourth line gives the degree of anisotropy $I$ (Eq.~(\ref{anisotropy})) of the fundamental optical transition with 
respect to the light polarization direction within the QD basal plane.
The fifth line shows the oscillator strength $f_{[001]}$ of the QD fundamental transition for light linearly
polarized along the [001] direction, normalized with respect to $f_{QD}$. 
The last line gives the relative change $(g-g_{0})/g_{0}$ of the optical QD gap $g$ with respect to the gap $g_{0}$ corresponding to the case ``strain off'' and ``interdiffusion off''. 
In the chemically disordered case, the error bar (when not negligible) was obtained by the same statistical analysis adopted in Fig.~\ref{strain_concentr_profiles}.  }
\end{table}

From the results in Table~\ref{results_table}, we arrive at the following conclusions:

\begin{enumerate}
\item The first two lines show that chemical disorder reduces the confinement of the 
QD bound states through the partial relief of the strain field, while the chemical effect does not directly contribute. 
In fact, chemical disorder reduces the charge fractions $\Delta Q_{|e1\rangle}$ and 
$\Delta Q_{|h1\rangle}$, while no changes are observed for the strain-unaffected calculation. 
The smaller confinement of the QD bound state wave functions in the chemically disordered case is consistent with the results of Fig.~\ref{energie_diff}, where all electron and hole bound states become shallower when chemical disorder increases.
\item The third line indicates that chemical disorder significantly enhances 
(by about $40\%$, in the case considered here) the oscillator strength 
$f_{QD}$ of the fundamental optical transition, in qualitative agreement with experimental
results.\cite{malik1987,leon1888}
This effect is primarily due to the modification of the  strain field due to chemical disorder, 
because in the strain-unaffected case $f_{QD}$ does not significantly vary.
\item From the fourth line, we observe that chemical disorder strongly reduces the in-plane asymmetry $I$,
in accordance with previous experimental results.\cite{ochoa192}
This is a direct consequence of the partial relief of strain due to disorder. 
In fact,\cite{santoprete_icps} in a pyramidal QD the asymmetry of the oscillator strength of the fundamental optical transition between the directions [110] and [1$\bar{1}$0]
is a direct consequence of asymmetry of the strain field between these directions, 
which is in turn a consequence of the $C_{2v}$ symmetry.
This can be deduced observing that in the strain-unaffected case $I$ vanishes.
This result could be experimentally exploited to detect, among different samples containing QDs of similar geometry, those
characterized by the higher chemical purity. 
In fact, these samples will be those with the higher asymmetry of the absorption 
coefficient of the fundamental optical transition (which is proportional to $I$), for in-plane polarized light.
\item The fifth line of Table~\ref{results_table} implies that chemical disorder weakens the fundamental 
optical transition for perpendicularly polarized light. 
This is a consequence of the strain relief inside the QD: In the limit of complete relief (QD strain-unaffected) 
this transition is strictly forbidden.\cite{santoprete_icps}
\item The last line summarizes the different effects contributing to
the blue shift in the fundamental optical transition with respect to a
hypothetical transition energy $g_0$ where both effects are removed. We
see that strain and chemical disorder increase the QD gap by the same
order of magnitude. 
We note that calculations for the relative blue shift presented in Ref.~\onlinecite{gunawan:05} systematically underestimate this quantity as compared to the experimental results in Ref.~\onlinecite{fafard2374} (see Fig. 7 in Ref.~\onlinecite{gunawan:05}). This discrepancy is probably due to the simplified theoretical description adopted there, where strain effects were not taken into account.
\end{enumerate}

Finally, we analyzed the $z$-component of the built-in dipole moment of the
electron-hole pair, and how disorder affects it. Such dipole moment experimentally shows up as a
Stark shift of the emitted light from a QD-LED under applied
electrical field.\cite{fry}
For pure pyramidal InAs/GaAs QDs, this dipole moment points towards
the base of the pyramid, i.e. the
center of mass of the electron ground state lies above that of the
hole ground state.\cite{stier}
However, in the case of truncated pyramidal In$_{x}$Ga$_{1-x}$As QDs,
with $x$ increasing from the base to the tip of the pyramid, the
dipole moment may have an opposite orientation, i.e. the center of mass of the hole state can sit
above that of the electron state.\cite{fry}
Some authors have argued that such inversion occurs also for QDs having
an In-rich core with an inverted-cone shape. This inverted-cone shape
has been observed in truncated-cone nominal In$_{0.8}$Ga$_{0.2}$As
QDs\cite{lenz5150} and In$_{0.5}$Ga$_{0.5}$As QDs.\cite{liu334}
In our case, the dipole moment is always directed towards the base of
the pyramid, i.e. the electron ground state sits always above the
hole ground state, both for the pure and the disordered QD. This is
because we have neither a truncated pyramidal shape nor an
In-concentration increasing from the base to the tip of the pyramid
(see Fig.~\ref{strain_concentr_profiles}).
However, we observe that the disorder decreases the dipole moment of
the dot. In fact, in the strained disordered case, the
center of mass of the electron state lies 2.8 \AA \ above that of the
hole state, while in the strained pure case this separation is 3.5
\AA.

\section{Summary and Conclusions}

We presented an atomistic interdiffusion model to simulate the composition profile of 
chemically disordered pyramidal In$_{x}$Ga$_{1-x}$As QDs buried in GaAs matrices.
Calculations for the strain field inside and around the disordered QDs were compared to the
strain field of chemically pure InAs QDs, showing that chemical disorder significantly reduces the absolute value of 
the strain field inside the QD, giving rise to smoother variations of this field across the heterointerfaces.
Furthermore, we analyzed the consequences of chemical disorder for the electronic and optical properties
within an ETB model. Our treatment allowed us to distinguish between two effects of the chemical disorder, namely the relief of the strain inside the QD, and the purely chemical effect due to the presence of new atomic species (Ga atoms) penetrating inside the QD.
We showed that these two components of disorder have comparable effects on the QD electronic spectrum, while for the optical properties the strain relief effects are more relevant.
In particular, we showed that strain relief (i) reduces the charge confinement (inside the QD) of
the electron and hole bound state wave functions, (ii) significantly enhances the oscillator strength of the fundamental optical transition, (iii) strongly reduces the asymmetry of the oscillator strength of the fundamental 
optical transition between the directions [110] and [1$\bar{1}$0] for in-plane polarized light, and (iv) strongly reduces the oscillator strength of the fundamental optical transition for perpendicularly polarized light.

Our results help to explain experimental findings for the optical properties of intermixed InAs/GaAs QDs.

\begin{acknowledgments}
This work was partially supported by the Brazilian agencies CNPq,
FAPERJ and Instituto do Mil\^{e}nio de Nanoci\^{e}ncias-MCT, and by
Deutsche Forschungsgemeinschaft within Sfb 296. 
BK thanks the hospitality of the CMTC at the University of Maryland.
\end{acknowledgments}

\bibliography{biblio}

\end{document}